\documentstyle[preprint,aps,floats,graphicx]{revtex}\begin{document}
\draft

\title{ Comment on ``Ferromagnetic film on a superconducting substrate''}
\author{E.B. Sonin}

\address{Racah Institute of Physics, Hebrew University of
Jerusalem, Jerusalem 91904, Israel \\ and \\ Ioffe Physical Technical
Institute, St.~Petersburg
194021, Russia}

\date{\today} \maketitle

\begin{abstract}

A superconducting substrate is not able to shrink drastically domains in a
ferromagnetic film, contrary to the prediction of Bulaevskii and Chudnovsky. This
is shown on the basis of the exact solution for the stripe domain structure. 
\end{abstract}


In Ref. \cite{BC}  Bulaevskii and Chudnovsky analyzed the equilibrium stripe domain
structure in a ferromagnetic film on a superconducting substrate and predicted a
drastic shrinkage of domains. According to them, the domain size is by the
factor $(\lambda_L/l)^{1/3}$ smaller than the  domain size $l\sim \sqrt{\delta d_M}$
for a film without a superconducting substrate \cite{LL}. Here $\lambda_L$ is the
London penetration depth, $d_M$ is the film thickness, and  $\delta$ is the domain
wall thickness ($\delta \ll l,~d_M$). In this Comment I shall show that this
prediction is incorrect: even in the limit $\lambda_L/l \rightarrow 0$,  the
superconducting substrate  can shrink domains only by a numerical factor not more
than $\sqrt{1.5}$.

\begin{figure}
  \begin{center}
    \leavevmode
    \includegraphics[width=0.9\linewidth]{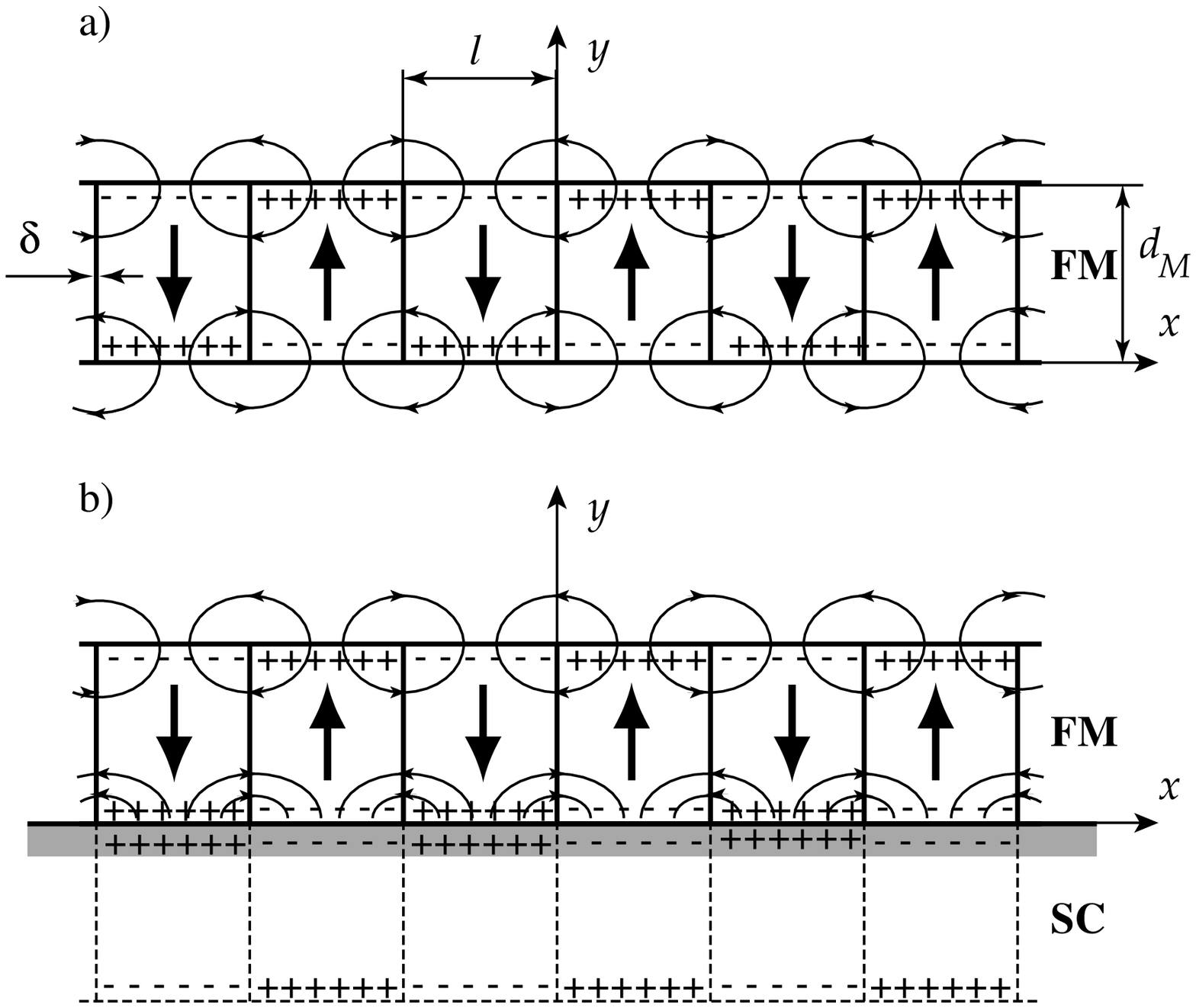}
    \bigskip
    \caption{Magnetic charges (+ and -) and magnetic flux (thin lines with arrows)    in a
ferromagnetic film (FM) without (a) and with (b) a superconducting substrate (SC).
The magnetization vectors in domains are shown by thick arrows.}
  \label{Fig}
  \end{center}
  \end{figure}

As well in Ref. \cite{BC}, I consider a ferromagnetic film with its 
spontaneous magnetic moment $\vec M$ normal to the film. The stray magnetic field
$\vec H=\vec B-4\pi \vec M$ must satisfy the equations of magnetostatics \cite{MC}:
\begin{equation}
\vec \nabla \times \vec H =0~,~~~\vec \nabla \cdot \vec H =4\pi \rho_M~,
 \label{ms} \end{equation}
where $\rho_M =-\vec \nabla \cdot \vec M$ is the magnetic charge, and $\vec M$ is the
spontaneous magnetization. The second equation in (\ref{ms}) follows from the
condition that the magnetic induction $\vec B= \vec H +4\pi \vec M$ is
divergence-free: $\vec \nabla \cdot \vec B=0$. If domain walls are 
parallel to  the magnetization $\vec M$, i.e., normal to the film, the magnetic
charges  appear only on the film surface (Fig. \ref{Fig}).

In the limit $\delta \ll l,~d_M$, which was considered in Ref. 
\cite{BC}, the distribution of stray fields for the stripe domain structure  
in a ferromagnetic film
can be found exactly using analytical functions on the complex plane
\cite{JTF}.  We assume that the film is parallel to the $xz$ plane and 
is restricted by the planes $y=0$ and $y=d_{M}$ (Fig. \ref{Fig}). The field
components $H_x$ and
$H_y$  satisfy Eqs. (\ref{ms}) if they are determined by the real and
imaginary parts of an analytical function ${\cal H}(w)$ on the complex plane
$w=x+iy$. Without a superconducting substrate the solution is
\begin{equation}
{\cal H}(w)=-H_x+ i H_y=4M \left[\ln\tan{\pi w\over 2l}-\ln\tan{\pi (w-id_M)\over
2l}\right]~.
 \label{sn} \end{equation}
If the film is put on a superconducting substrate with the London penetration depth
much less than the domain size $l$ and the film thickness $d_M$ (the  case when
Bulaevskii and Chudnovsky predicted a strong effect of the substrate), one can
neglect the penetration of the magnetic field into the substrate, and obtain the
solution of the problem by introducing image charges in the substrate:
\begin{equation}
{\cal H}(w)=-H_x+i H_y=4M \left[2\ln\tan{\pi w\over 2l} 
-\ln\tan{\pi (w-id_M)\over
2l}-\ln\tan{\pi (w+id_M)\over 2l}\right]~.
 \label{ss} \end{equation}
The solutions Eqs. (\ref{sn}) and (\ref{ss}) are a straightforward generalization of
the solutions for a single domain wall  obtained in Ref. \cite{JTF}. The single-wall
solutions ($l\rightarrow \infty$) of Ref. \cite{JTF} follow from Eqs. (\ref{sn}) and
(\ref{ss}) after expansion of the tangent function: $\tan \varphi \approx \varphi$.

Later on we restrict ourselves to the case when the film thickness $d_M$
essentially exceeds the domain structure period $l$. Then the stray fields on two
film boundaries ($y=0$ and $y=d_M$) do not overlap and can be calculated
separately.  Near the boundary $y=0$ in absence of a superconducting substrate
\begin{equation}
{\cal H}(w)=4M \left(\ln\tan{\pi w\over 2l}-i{\pi \over 2} \right)~.
 \label{snS} \end{equation}
In presence of a superconducting substrate Eq. (\ref{ss}) yields by a factor 2 larger 
values of ${\cal H}$ at $y>0$, but ${\cal H}=0$  at $y<0$. However, one should
remember that we are solving the problem in the limit $\lambda_L \rightarrow 0$.
For finite $\lambda_L$ the jump of the tangential component $H_x$ at the plane
$y=0$ transforms into an exponential decrease of $H_x$ at $y<0$ down to zero at the
distance $\lambda_L$, and $H_x$ is continuous in accordance with the laws of
electrodynamics (see below).

Especially important for us is the magnetic field at the ferromagnetic film 
boundary $y=0^+$. Without a superconducting substrate:
\begin{equation}
H_{x}(x)=- \mbox{Re}{\cal H}=-4M \ln\left|\tan{\pi x \over 2l}\right|~.
 \label{x0} \end{equation}
\begin{equation}
H_{y}(x)= \mbox{Im}{\cal H}=\mp 2\pi M \mbox{sign} \left(\tan{\pi x \over 2l}\right)
~~~~
\mbox{at}~  ~y \rightarrow \pm 0~.
 \label{y0} \end{equation}
The  field pattern is periodic with the period $2l$ along the axis $x$. The magnetic
charge on the film boundary 
 $y=0$ is 
\begin{equation}
\rho_{M}={1\over 4\pi}[H(x+i0)-H(x-i0)] \delta(y)
=- M  \delta(y)
\mbox{sign} \left(\tan{\pi x \over 2l}\right)~.
    \label{ch} \end{equation}

So in the limit of $\lambda_L \rightarrow 0$ the method of complex variables
provides the exact solution of the problem in terms of elementary functions without
using the Fourier expansion. For finite $\lambda_L$ the exact solution in the form
of the infinite Fourier series is also known and agrees with our $\lambda_L
\rightarrow 0$ solution. One may check it comparing the magnetic energy of the two
solutions.  The magnetic energy can be calculated using the potential $\phi$ for 
the magnetic field ($\vec H=\vec \nabla \phi$) and integration by parts:
\begin{equation}
{\cal E}_{m}=\int d V {H^{2}\over 8\pi}={1\over 2} \int d S \rho_M \phi  ~,
 \label{E} \end{equation}
where the surface integral should be taken over all
planes which confine the magnetic charge $\rho_{M}$. Without a superconducting
substrate in the limit $d_M \gg l$ the energy of  the stray fields near the plane
$y=0$ per unit area in the plane $xz$ is: 
\begin{eqnarray}
E_{m}={ M\over 2 l}\int _{0}^{l} dx \int _{0}^{x} dx'4M \ln\tan{\pi x' \over
2l}={8 M^{2} l \over \pi^2} \int _{0}^{\pi \over 2}d\varphi\, \varphi\ln\tan{\varphi
}  \nonumber \\
={7 \zeta(3) \over \pi^2}  M^{2} l\approx 0.852  M^{2} l~,
 \label{E-d} \end{eqnarray}
where $\zeta(z)$ is the zeta function. The same value of energy was obtained with
the Fourier--expansion method in the problem after Sec. 44 in the book by Landau and
Lifshitz \cite{LL}. The Fourier--series solution for the magnetostatic problem of a
ferromagnetic film on a superconducting substrate for arbitrary $\lambda_L$ was
found by Stankiewicz {\em et al} \cite{ST}, and their solution also agrees with  our 
$\lambda_L \rightarrow 0$ solution. This is checked in details in Ref. \cite{LLST}.
 
The superconducting substrate increases the magnetic energy density in the
film by four times, but contracts the area occupied by the magnetic field 
by two times. Thus the magnetic energy at the boundary $y=0$  in presence of the
substrate is by two times  larger than without it. On the other hand, in our limit
$d_M \gg l$ the substrate has no effect on the magnetic energy at the other boundary
$y=d_M$. Eventually the substrate increases the total magnetic energy by 1.5 times.
The energy of the domain walls per unit length along the axis $x$ is inversely
proportional to the period $l$ and the energy of the stray fields is proportional to
$l$. The period
$l$ is determined by minimization of the total energy per unit length, and the growth
of the magnetic energy by two times decreases the domain width $l$ only by
$\sqrt{1.5}$ times. 

In any domain the surface charges 
on the film boundary $y=0$ generate the magnetic flux $\Phi=\pm 4\pi M l$. Without a 
superconducting substrate, half of this flux enters the film itself, 
and another half exits from the film (Fig. \ref{Fig}a). The superconducting substrate 
does not allow for the magnetic flux to exit from the film, and the 
whole flux enters the film (Fig. \ref{Fig}b). Let us consider now the effect of a
small,  but finite London penetration depth. The magnetic field inside the 
superconductor is determined by the boundary value of the tangential field $H_x$ in 
the ferromagnetic film at $y =0$, which is of the order of $M$. Then the
magnetic  flux, which enters the superconductor, is  $\sim  M \lambda_{L}$, {\em
i.e.},   about $\lambda_{L}/l$ times smaller than the total flux $4\pi M l$. 
This provides a correction of the relative order $\lambda_{L}/l$ to the 
magnetic flux of the stray magnetic fields inside the film. The
energy $\sim M^2 \lambda_L$ inside the superconductor is also a small
correction of the same relative order.

The latter discussion helps to understand the source of an error in 
Ref. \cite{BC}. Looking for the magnetic field distribution, Bulaevskii and 
Chudnovsky \cite{BC} assumed that the magnetic field component normal to the film 
boundary is the same inside the film and inside the superconducting  substrate [see
their Eq. (7)] . So according to their solution half  of the total stray magnetic
flux enters the superconductor even in the  limit $\lambda_{L} \rightarrow 0$.
Meanwhile only a small part $\propto \lambda_{L}/l$ of the total flux  is able to
penetrate to the superconducting substrate. It is worth to stress that the solution
of the problem does not need any {\em apriori} assumption on distribution of the
magnetic flux between the ferromagnet and the superconductor at all. One must simply
use correct electrodynamic boundary conditions \cite{LL} at the interface $y=0$: 
continuity of the normal component of the magnetic induction $\vec B$ and continuity
of the tangential component of the magnetic field $\vec H$. The solution by
Bulaevskii and  Chudnovsky satisfies the first condition but violates the second
one. Indeed, the values of the Fourier components of $H_x$ inside and outside the
superconductor, which are  given by Bulaevskii and  Chudnovsky after their Eq. (8),
differ by a large factor $1/q \lambda_L$, where $q \sim 1/l$ is the wave number in
the Fourier expansion used by  Bulaevskii and  Chudnovsky. Because of  this error,
they essentially overestimated the energy inside the
superconductor, and as a result of it,  predicted a strong shrinkage of the domains. 

In summary, the result of Bulaevskii and  Chudnovsky \cite{BC} on domain structure
in a ferromagnetic film on a superconducting substrate is incorrect because they
ignored the electrodynamic boundary condition that the tangential component of the
magnetic field must be continuous at the ferromagnet--superconductor interface.
Instead of it they used the incorrect assumption that the magnetic flux produced by
magnetic charges at the interface is equally distributed between  the
ferromagnet and superconductor. The correct solution of the problem in terms of
elementary analytic functions on the complex plane is given, which is exact in the
limit of large ratio of the domain size to the London penetration depth.

I acknowledge helpful discussions and comments by N.B. Kopnin, K.B. Traito, G.E.
Volovik, M. Ziese, and especially E.H. Brandt.  The work was supported by the grant
of the Israel Academy of Sciences and Humanities.

\newpage

\end{document}